\documentclass[twoside,leqno]{article}

\usepackage[letterpaper]{geometry}
\usepackage{lipsum}
\usepackage{amsfonts}
\usepackage{ltexpprt}
\usepackage{hyperref}
\usepackage{graphicx}
\usepackage{epstopdf}
\usepackage{amsmath}
\usepackage{xcolor}
\usepackage{url}
\usepackage{algorithmic}

\ifpdf
\hypersetup{
  pdftitle={TopoEmbedding},
  pdfauthor={X. Bao, G. Liu, and F. Iuricich}
}
\fi

\begin{document}

\newcommand\relatedversion{}

\title{TopoEmbedding, a web tool for the interactive analysis of persistent homology}


\author{Xueyi Bao\thanks{Clemson University, Clemson, SC (\{xueyib, guoxil, fiurici\}@g.clemson.edu).}
\and Guoxi Liu\footnotemark[1]
\and Federico Iuricich\footnotemark[1]}

\date{}

\maketitle


\fancyfoot[R]{\scriptsize{Copyright \textcopyright\ 2022 by SIAM\\
Unauthorized reproduction of this article is prohibited}}





\begin{abstract} \small\baselineskip=9pt Software libraries for Topological Data Analysis (TDA) offer limited support for interactive visualization. Most libraries only allow to visualize topological descriptors (e.g., persistence diagrams), and lose the connection with the original domain of data. This makes it challenging for users to interpret the results of a TDA pipeline in an exploratory context. In this paper, we present TopoEmbedding, a web-based tool that simplifies the interactive visualization and analysis of persistence-based descriptors. TopoEmbedding allows non-experts in TDA to explore similarities and differences found by TDA descriptors with simple yet effective visualization techniques.\end{abstract}

\section{Introduction.}
Topological Data Analysis (TDA) has successfully demonstrated the ability to address the needs of data scientists when studying various data types: from scalar fields \cite{Heine2016}, to high dimensional point clouds \cite{Singh2007}. One advantage of TDA is that it provides compact descriptors describing essential features of the data. 
Many TDA software libraries are available nowadays, implemented in different programming languages, such as R, Python, and C++. These include Javaplex \cite{hong_javaplex_2014}, RedHom \cite{hong_capdredhom_2014}, Gudhi \cite{hong_gudhi_2014}, Scikit-tda \cite{scikittda2019}, PHAT \cite{mcgeoch_distributed_2014}, DIPHA \cite{bauer_phat_2017}, Dionysus \cite{morozov2007dionysus}, Perseus \cite{mischaikow_morse_2013}, Ripser \cite{marfil_persistent_2019}, HomCloud \cite{HomCloud}, and Diamorse \cite{delgado2015diamorse}. All these libraries dedicate particular attention to persistent homology, one of the most common tools in TDA.

{\em Persistent homology} \cite{Edelsbrunner2002,zomorodian_computing_2005} provides a multiscale description of the homology of a shape by means of a filtration. Let $\Sigma$ be a simplicial complex, a {\em filtration} is a sequence $\{\Sigma^i\, |\, 0\leq i \leq r\}$ of subsets of $\Sigma$ such that $\emptyset=\Sigma^0 \subseteq \Sigma^1 \subseteq \dots \subseteq \Sigma^r=\Sigma$. Intuitively, given an index $i$, homology captures the cycles of $\Sigma^i$  \cite{munkres_elements_2018}. Given two indices $i$ and $j$, with $i < j$, persistent homology tracks the cycles that appear, or disappear, from $\Sigma^i$ to $\Sigma^j$. Each cycle is characterized by a pair of indices $(i,j)$,  where $i$ is called {\em birth}, $j$ is called {\em death}, and $j-i$ is called {\em persistence}. Each pair is also referred to as a {\em persistence pair}, and the collection of all pairs defines the {\em persistence diagram} \cite{cohen-steiner_stability_2007}. 

To simplify the analysis of persistence information, most libraries visualize the persistence diagram as a scatter plot representing each persistence pair as a point with the coordinate value $(i,j)$. An alternative representation is a barcode that represents each feature as a line \cite{ghrist_barcodes_2007}. Two libraries offer alternative ways to visualize persistence. Eireen \cite{henselman_matroid_2016} is a toolbox implemented in Julia \cite{bezanson_julia_2017} for the analysis of Vietoris-Rips filtrations \cite{quinn_prospects_1996}. On top of the persistence diagram, it offers the explicit computation and visualization of the associated persistence cycles. The Topological ToolKit (TTK) \cite{tierny_topology_2018} focuses on topological data analysis for scientific visualization and defines a unified framework for visualizing different topological structures such as Reeb graphs \cite{reeb1946points}, contour trees \cite{van1997contour}, and Morse-Smale complexes \cite{milnor_morse_1963}. It also provides two ways to visualize persistent homology other than the persistence diagram: the embedded persistence pairs, which represent each dot of the persistence diagram as a pair of points in the filtration domain, and the persistence curve, which is a line chart plotting the number of pairs at the vary of the persistence values. Recently, a new plugin for TTK has been developed which allows the visualization of persistence cycles \cite{Iuricich2021}, an explicit representation of the boundary of each hole, or void originated by the filtration.\\

Rather than visualizing information related to a single descriptor, this manuscript considers the problem of analyzing multiple descriptors at once. While many metrics have been studied to measure the similarities and differences among persistence-based descriptors \cite{Yan2021}, results are often presented in a "black-box" manner. That is, measures are used to quantify what descriptors are the most similar without explaining why. In an effort to simplify the interpretation of TDA descriptors for non-experts, this paper proposes TopoEmbedding, an interactive web tool for the analysis and explanation of persistence features similarities.

\section{Overview.}
\label{sec:motivation}
The goal of TopoEmbedding is to simplify the analysis of topological descriptors by means of interactive visualizations. In Section \ref{sec:analysis} we describe the TDA pipeline we used to compute the persistence-based descriptors rendered by TopoEmbedding. In Section \ref{sec:visualization} we describe the visual interface used to improve explainability and interactive analysis.

\subsection{Analysis pipeline.}
\label{sec:analysis}
To demonstrate the interface functionality, we implemented a representative TDA pipeline analyzing the famous MNIST data set, a collection of 1000 handwritten digits (100 images per digit). The analysis pipeline consists of three different stages. 

In the first stage, we compute persistent homology for each handwritten image. If we consider each input image as a piecewise linear scalar field $f: \Sigma \rightarrow \mathbb{R}$ defined on a simplicial complex $\Sigma$, a filtration of $\Sigma$ is naturally defined by the sequence of sublevel sets according to $f$, where the {\em sublevel set} of value $i$ is defined as $\Sigma^i = \{ \sigma \in \Sigma | f(\sigma) \leq i \}$. Then, given the filtration $f$ defined on each input image we can obtain the corresponding persistence diagram using the standard algorithm \cite{Edelsbrunner2002}. In this analysis, we focus on points of the persistence diagram representing 1-cycles only and we discard the remaining ones.\\

During the second stage, we generate a persistence image \cite{adams2017persistence} for each persistence diagram computed in the first stage. Intuitively, a {\em persistence image} is a regular grid discretizing the density of persistence pairs in each portion of the persistence diagram. Each persistence pair $(i, j)$ in a persistence diagram is first transformed into a birth-death coordinates representation with $(x, y) = (i, j-i)$. Then, a persistence image is obtained by fitting a Gaussian kernel distribution upon each persistence pair $(x, y)$ defined as:
$$ g(x, y) = \frac{1}{2\pi \sigma^2}e^{-[(x-u_x)^2 + (y-u_y)^2]/2\sigma^2} $$
with mean $u = (u_x, u_y)$ and $\sigma = 0.01$. Each distribution is sampled on a $n=10\times 10$ grid and weighted by the function 
\begin{equation*}\label{e1.1}
    w_b(y) = 
    \begin{cases}
        0, & \text{if } y \leq 0 \\
        \frac{y}{b}, & \text{if } 0 < y < b \\
        1, & \text{if } y \geq b, 
    \end{cases}
\end{equation*}
where $b$ is the persistence value of the most persistent feature.\\



The idea of the third stage is to interpret the input dataset as a higher dimensional point cloud. After computing the persistence image for each input dataset, we compute a matrix of pairwise distances for all of them with the 2-norm Minkowski distance, which is defined as $D(X, Y) = \sqrt{\sum^{n}_{i=1}|p_i - q_i|^2}$, where $X$ and $Y$ denote two persistence images, $p_i$ and $q_i$ are the $i$-th pixels from persistence image $X$ and $Y$, respectively.

Notice we use persistent images because of their robustness and scalability \cite{adams2017persistence}. However, the metrics $D$ could be computed directly on the persistent diagram using bottleneck distance \cite{cohen-steiner_stability_2007}, Wasserstein distance \cite{cohen-steiner_lipschitz_2010}, or similar metrics \cite{yan_scalar_2021}.
 We use the matrix of pairwise distances to obtain a lower-dimensional embedding of the point cloud, where each point corresponds to an input digit image (associated with a persistence diagram and also a persistence image).

A number of supervised and unsupervised dimensionality reduction techniques are easily available to the scope. In our analysis we use three methods simultaneously, namely, Isomap\cite{balasubramanian_isomap_2002}, Multi-dimensional Scaling (MDS)\cite{webb_multidimensional_1995}, and t-distributed Stochastic Neighbor Embedding (t-SNE)\cite{van2008visualizing}. Figure \ref{fig:embedding} shows an example of the embedding obtained with Isomap.

The whole process is computationally efficient and scalable. All our experiments were run on a Desktop computer mounting a Core i7-8700 3.20 GHz processor and 32 GB of RAM. In the first stage we used a dedicated TTK plugin for computing persistent homology and persistence cycles on each image \cite{Iuricich2021}. This step requires less than 50MB of RAM and less than a second per image. In the second and third stages, the peak memory usage for computing the persistent images and the lower dimension embedding (including the distance matrix) is less than 100MB, and the computation time is around 120 seconds. All computations are performed offline and the results saved in dedicated files used during the interactive phase.

\subsection{Visualization and user interactions.}
\label{sec:visualization}

\begin{figure}[htp]
   \centering
   \includegraphics[width=0.8\linewidth]{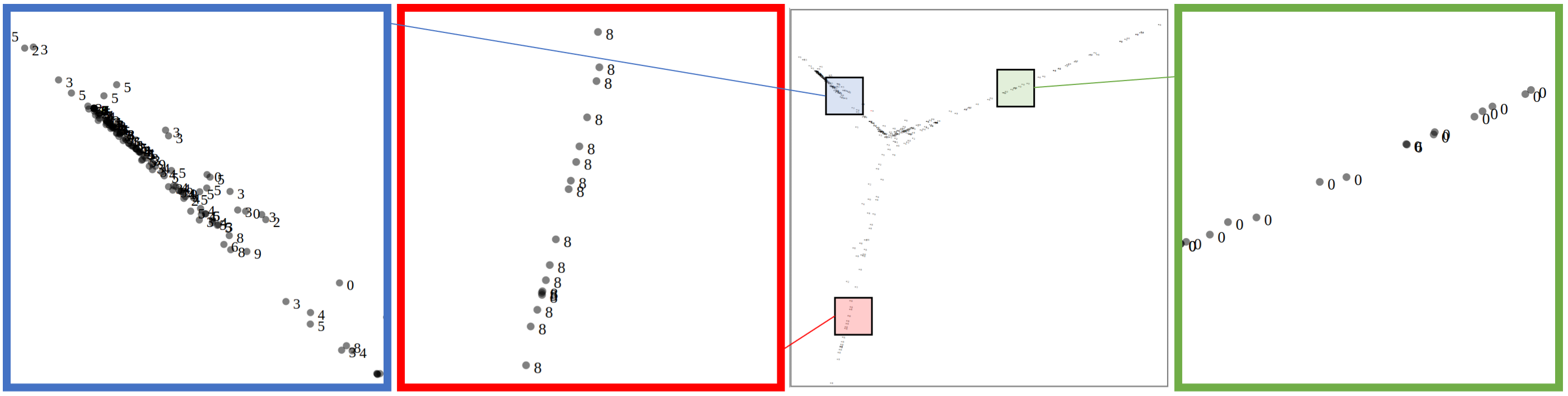}
   \caption{Three zoomed views of the lower dimensional embedding created by Isomap. Each view is color-coded to indicate where it was sampled from. We can notice that each branch of the Isomap embedding isolates digits with zero highly persistent 1-cycles (blue), two 1-cycles (red), or one 1-cycle (green).}
   \label{fig:embedding}
\end{figure}

The lower-dimensional embedding obtained with standard data science libraries \cite{balasubramanian_isomap_2002,webb_multidimensional_1995,van2008visualizing} is directly visualized as a 2D scatterplot.
Figure \ref{fig:embedding} shows the results obtained with Isomap \cite{balasubramanian_isomap_2002}. We can notice that Isomap is preserving the information about the input digit similarities. We recall that the analysis pipeline focuses on 1-cycles (see Section \ref{sec:analysis}). Thus, it is not surprising that the lower-dimensional embedding highlights three classes of similarity among the input digits: digits with one distinct 1-cycle (e.g., 6,9,0), digits with two distinct 1-cycles (e.g., 8), and digits with no 1-cycles (e.g., 1,2,3,4,5,7).

\begin{figure}[htb]
   \centering
   \includegraphics[width=0.8\linewidth]{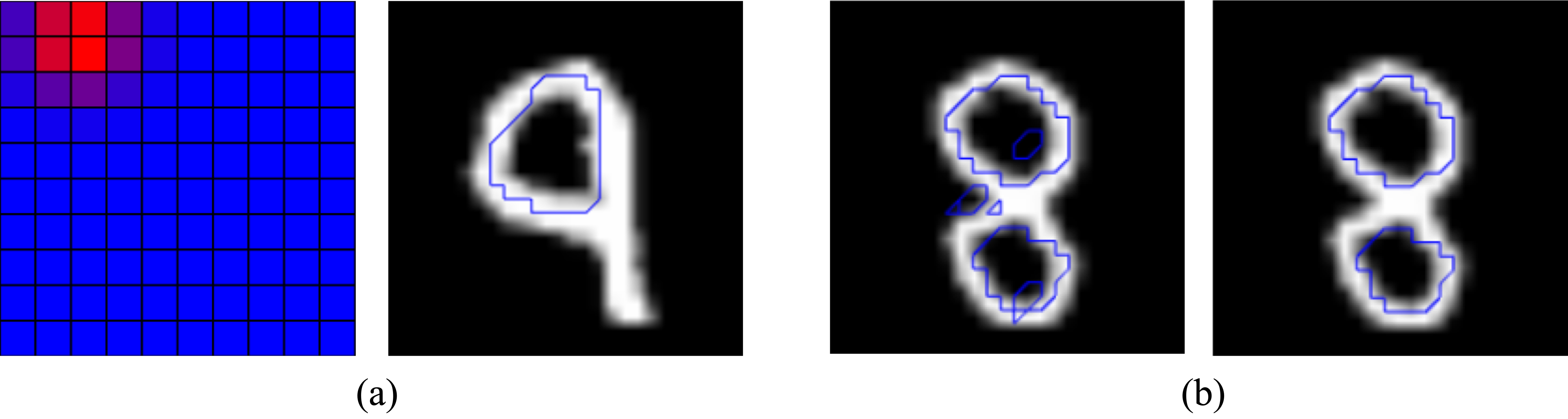}
   \caption{(a) Example of a persistence image and the corresponding persistence cycles as rendered in TopoEmbedding. The persistence image is visualized as a grid of resolution $n=10\times10$ and with pixel values color-coded according to a diverging (blue-red) color map. The 1-cycles (right) are displayed upon the digit as collections of blue lines indicating the boundary of a hole originated by the filtration. (b) Cycles can be filtered according to their persistence. This allows to update the visualization focusing on the most important features captured by persistent homology. On the left we show unfiltered cycles as computed by persistent homology. On the right we show only the most persistent cycles.}
   \label{fig:persistence}
\end{figure}

It is fairly easy for a user who is experienced in TDA and persistent homology to interpret the representation. However, what about an inexperienced one? To address this need, TopoEmbedding allows any user to visually explore the lower-dimensional embedding and its corresponding persistence images along with the 1-cycles. When the user selects a point in the scatter plot, TopoEmbedding will display the corresponding persistence image (see Figure \ref{fig:persistence}a left) in a diverging blue-red color map. Red pixels indicate high-density regions that contain more persistence pairs than the blue pixels. The visualization of the persistence image is also paired with the input digit image, and the persistence cycles \cite{Iuricich2021} (see Figure \ref{fig:persistence}a right), an explicit visualization of 1-cycles originated and destroyed by the filtration. We recall that persistent homology associates an importance value (i.e., persistence) to each feature according to its lifespan in the filtration. Using this information, we are able to provide a filtering tool allowing users to remove cycles according to their importance (see Figure \ref{fig:persistence}b).
\\

\begin{figure}[htb]
   \centering
   \includegraphics[width=0.8\linewidth]{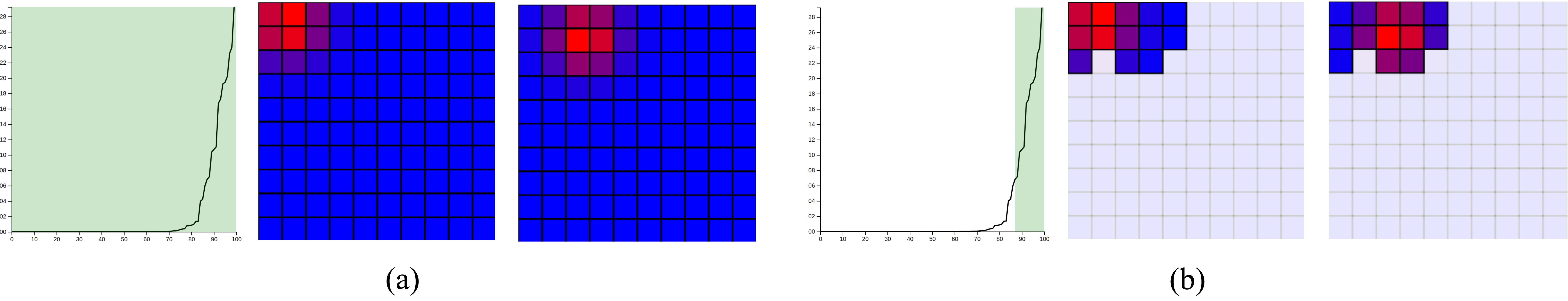}
   \caption{(a) Line chart showing the pixel-wise difference between the two persistence images selected. The green background in the line chart indicates the pixel-wise difference selected as threshold. (b) After reducing the thresholding area focusing on pixels with the highest pixel-wise difference, the green area in the chart is resized as well as all pixels in the persistence images excluded by the same threshold are displayed with higher transparency colors.} 
   \label{fig:pixel_threshold}
\end{figure}

By selecting two digits, the user can compare their corresponding descriptors and topological features. The two persistence images are juxtaposed, and a line chart (see Figure \ref{fig:pixel_threshold}a) shows the pixel-wise difference between two persistence images with values sorted in ascending order. 
To highlight differences between the two persistence images, the user can threshold their pixels setting the minimum and maximum pixel-wise difference. As a result, only pixels within such range will be visualized at full opacity (see Figure \ref{fig:pixel_threshold}b).
\\

The interface components can be used to explore similarities and differences between a pair of persistence images. At the same time, differences in the corresponding topological features (i.e., persistence cycles) visualized in the input data domain space explain the causes of such differences in an interpretable manner.

\begin{figure}[htp]
   \centering
   \includegraphics[width=0.8\linewidth]{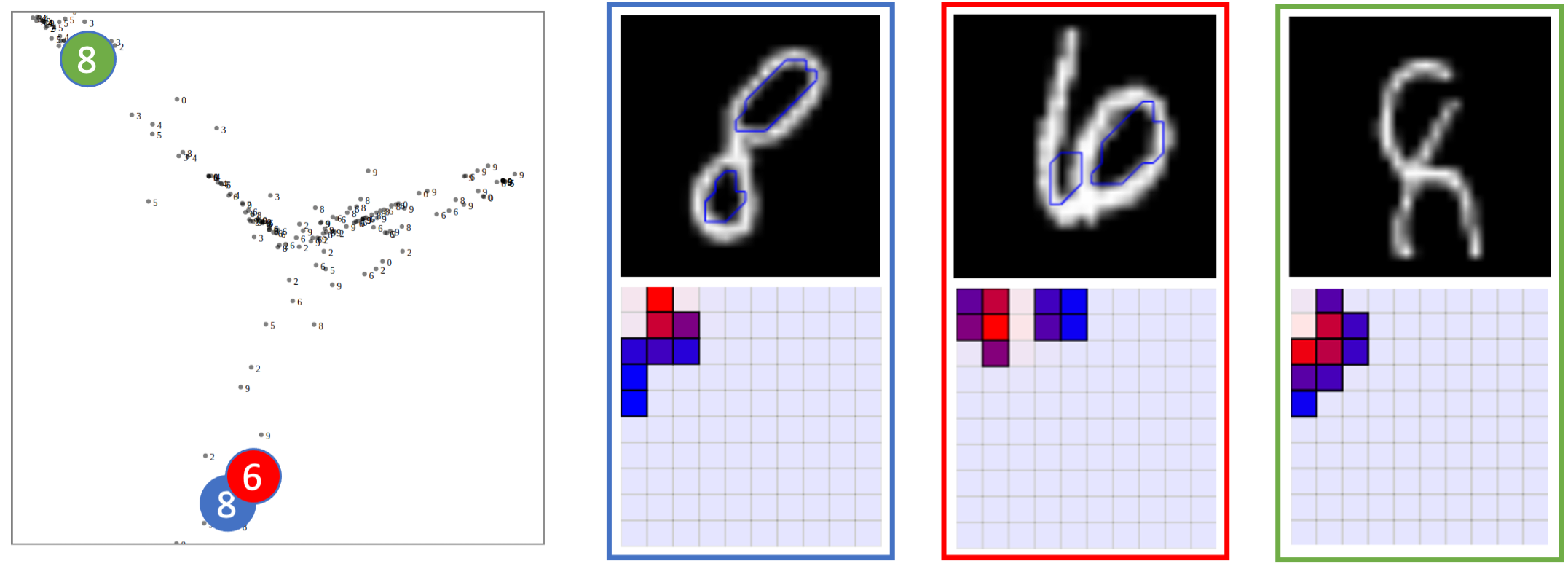}
   \caption{While the digit 8 highlighted in blue belongs to the correct branch where most digits 8 are, a few digits in the MNIST dataset are being classified as belonging to the wrong branch of the Isomap embedding. One of these is digit 6 highlighted in red. While the corresponding persistence image does not clarify why the digit is classified similarly to an 8, the explicit visualization of the 1-cycles shows that the underlying filtration created two highly persistent 1-cycles. The other way around, the digit 8 highlighted in green ends up far away from its corresponding branch. Once again, the visualization of the 1-cycles shows that there are not highly persistent 1-cycles in this filtration, explaining why the digit ends up in said position.}
   \label{fig:similarity}
\end{figure}

As an example, Figure \ref{fig:similarity} illustrates a possible use case specific to the MNIST dataset. The digit 8 highlighted in blue in Figure \ref{fig:similarity} correctly belongs to the branch of the Isomap embedding containing most 8 digits (i.e., those containing two cycles). 
\\

Close to such digit, we found digit 6 highlighted in red. This digit should not belong to this branch since it should be characterized by one 1-cycle. If we compare the most different pixels in the corresponding persistence images, it is still difficult to interpret why one digit is being classified as close to the other. However, if we look at the explicit representation of 1-cycles (Figure \ref{fig:similarity}), we notice that both filtrations create two high persistent 1-cycles. This explains why digit 6 is being classified similarly to digit 8.

The same reason can be applied to a digit 8 (highlighted in green), which ends far away in the embedding from the first digit 8 (highlighted in blue). Once again, the corresponding persistence image is hard to interpret for humans. However, if we combine the persistence image with an explicit representation of the persistence cycles, we can realize that no highly persistent cycles are being captured for this digit. This explains why this digit is classified as having no cycles and, for this reason, is being correctly grouped with similar digits (e.g., 1,2,3, etc.)

\section{Conclusions}
\label{sec:conclusions}

We have presented TopoEmbedding, a web-based TDA tool to assist in the visualization and analysis of persistent homology. The source code of our analysis pipeline and the implementation of TopoEmbedding can be found at \url{https://github.com/DaVisLab/TopoEmbedding}.
TopoEmbedding represents a proof-of-concept highlighting the importance of interactive visualizations for TDA. In this work, we used a 2D dataset (e.g., MINIST) to demonstrate the functionality of TopoEmbedding. However, we are interested in applying similar strategies for the analysis of 3D datasets where the direct rendering of volumetric images is unfeasible. In such a context, the visualization of 1-cycles and 2-cycles (surfaces indicating the boundary of voids) could become crucial for interpreting the results of the TDA pipeline.




\bibliographystyle{siamplain}
\bibliography{main}
\end{document}